\documentstyle[12pt,epsfig]{article}
\def\bge{\begin{equation}}
\def\ene{\end{equation}}
\def\bg{\begin{eqnarray}}
\def\en{\end{eqnarray}}
\def\non{\nonumber}

\def\Dbar{\overline{D}}
\def\D0bar{\overline{D^0}}

\begin{document}
\begin{titlepage}
\title{$\Lambda^+_c$- and $\Lambda_b$-hypernuclei 
}
\author{
K. Tsushima$^1$~\thanks{tsushima@physast.uga.edu}~,
F.C. Khanna$^2$~\thanks{Permanent address: 
Department of Physics, University of Alberta, Edmonton, Canada, T6G 2J1
\newline\hspace*{1.5em} khanna@Phys.UAlberta.CA}
\\ \\
{$^1$\small Department of Physics and Astronomy,
University of Georgia, Athens, GA 30602, USA} \\
{$^2$\small CSSM, University of Adelaide, Adelaide, SA 5005, Australia}
}
\maketitle
\vspace{-8cm}
\hfill Alberta Thy 12-02
\vspace{8cm}
\begin{abstract}
$\Lambda^+_c$- and $\Lambda_b$-hypernuclei are studied 
in the quark-meson coupling (QMC) model. Comparisons are made with 
the results for $\Lambda$-hypernuclei studied in the same model previously.
Although the scalar and vector potentials felt by the $\Lambda$, 
$\Lambda_c^+$ and $\Lambda_b$ in the corresponding hypernuclei multiplet 
which has the same baryon numbers are quite similar,  
the wave functions obtained, e.g., for $1s_{1/2}$ state, are very different.
The $\Lambda^+_c$ baryon density distribution in $^{209}_{\Lambda^+_c}$Pb 
is much more pushed away from the center 
than that for the $\Lambda$ in $^{209}_\Lambda$Pb due to the Coulomb force.
On the contrary, the $\Lambda_b$ baryon density distributions 
in $\Lambda_b$-hypernuclei are much larger near the origin than those for 
the $\Lambda$ in the corresponding $\Lambda$-hypernuclei 
due to its heavy mass. 
It is also found that level spacing for the $\Lambda_b$ 
single-particle energies is much smaller than that for   
the $\Lambda$ and $\Lambda^+_c$. 
\\ \\
{\it PACS number(s)}: 24.85.+p, 14.20.Lq, 14.20.Mr, 21.80.+a, 21.65.+f  
\end{abstract}
\end{titlepage}
%
Recently we have made  
a systematic study of the changes in properties of the  
heavy hadrons which contain a charm or a bottom quark 
in nuclear matter~\cite{TK}.
The results suggest that the formations of 
charmed and bottom hypernuclei, 
which were predicted first in mid 70's~\cite{Tyapkin,Dover}, 
are quite likely.  
The experimental possibilities were also studied later~\cite{bcexp}.
In addition, we predicted the $B^-$-nuclear (atomic) 
bound states, based on analogy with kaonic atom~\cite{Katom}, and 
a study made for the $D$- and $\Dbar$-nuclear 
bound states~\cite{Tsushimad} using the quark-meson coupling (QMC) 
model~\cite{Guichon,Guichonf,Saitof}. 

The QMC model, which was used there and is used  
in this study, has been successfully 
applied to many problems associated with nuclear physics and hadronic 
properties in nuclear medium~\cite{Tsushima_hyp,QMCapp}. 
For example, the model was applied to the study of 
$J/\Psi$ dissociation in nuclear matter~\cite{Alexjpsi}, and 
$D$ and $\Dbar$ productions in antiproton-nucleus
collisions~\cite{Alexd}.
Furthermore, although only limited studies for heavy mesons 
(not for heavy baryons) with charm
in nuclear matter were made by QCD sum rule 
(for $J/\Psi$~\cite{Hayashigaki1,Klingl} and 
$D (\Dbar)$~\cite{Hayashigaki2}), 
a study~\cite{TK} for heavy baryons with a charm
or a bottom quark based on quarks was made using QMC.
In particular, recent measurements of polarization transfer performed
at MAMI and Jlab~\cite{JlabQMC} support the medium modification of the
proton electromagnetic form factors calculated by the QMC model.
The final analysis~\cite{Strauch} 
seems to become more in favor of QMC, although   
still error bars may be large to draw a definite conclusion.

Certainly, the model has shortcomings to be improved eventually. 
Difficulties to handle it will be increased rapidly  
if we adopt Hartree-Fock approximation even for nuclear 
matter~\cite{QMCfock}, and the inclusion of Pauli blocking  
at the quark level, and the $\Sigma N - \Lambda N$ channel 
coupling have not been implemented yet in a consistent manner with 
the underlying quark degrees of freedom~\cite{Tsushima_hyp}.
(It should be mentioned that in the case of $\Sigma$-hypernuclei 
no narrow states have been observed. It is unlikely that 
it will be possible to find such states 
in the present case~\cite{sigmahyp}.)
Furthermore, an application to double hypernuclei has not 
been attempted, although recently the existence
was confirmed~\cite{doublehyp}.
Nevertheless, with its simplicity and 
successful applicability achieved so far, 
we feel some confidence that such a quark-meson coupling model will provide
us with valuable glimpse into the
properties of charmed- and bottom-hypernuclei.

In this article, we make a quantitative study for the $\Lambda^+_c$- 
and $\Lambda_b$-hypernuclei, by solving a system of 
equations for finite nuclei,  
embedding a $\Lambda^+_c$ or a $\Lambda_b$ to the closed-shell nucleus
in Hartree, mean-field, approximation.
Then, the results are compared with those for the 
$\Lambda$-hypernuclei~\cite{Tsushima_hyp}, 
which were studied in QMC. 
It is shown that, although the scalar and vector potentials 
felt by the $\Lambda$, $\Lambda^+_c$ and $\Lambda_b$ 
in the corresponding hypernuclei multiplet which has the same baryon numbers 
are quite similar, the wave functions obtained, e.g., for  
$1s_{1/2}$ state are very different. Namely,  
the $\Lambda^+_c$ baryon density distribution in $^{209}_{\Lambda^+_c}$Pb 
is much more pushed away from the center than that
for the $\Lambda$ in $^{209}_\Lambda$Pb due to the Coulomb force.
On the contrary, the $\Lambda_b$ baryon density distributions 
in $\Lambda_b$-hypernuclei are much more central 
than those for the $\Lambda$ in the corresponding $\Lambda$-hypernuclei 
due to its heavy mass.
In addition it turns out that the level spacing for the 
$\Lambda_b$ single-particle energies is much smaller than that for 
the $\Lambda$ and $\Lambda^+_c$, which may imply many interesting 
new phenomena, which will be discovered in due course 
by experiments. We hope this study opens 
a new possibility for experiments, related to nuclear and hadronic 
physics, especially for 
Japan Hadron Facility (JHF). 

We start to consider static, (approximately) spherically symmetric
charmed and bottom hypernuclei (closed shell plus one heavy baryon
configuration) ignoring small non-spherical 
effects due to the embedded heavy baryon.
We adopt Hartree, mean-field, approximation.
In this approximation, $\rho NN$ tensor coupling gives
a spin-orbit force for a nucleon bound 
in a static spherical nucleus, although
in Hartree-Fock it can give a central force which contributes to
the bulk symmetry energy~\cite{Guichonf,Saitof}. 
Furthermore, it gives no contribution for nuclear
matter since the meson fields are independent of position
and time. Thus, we ignore the $\rho NN$ tensor coupling in this
study as usually adopted in the Hartree treatment of
quantum hadrodynamics (QHD)~\cite{QHD}.

Using the Born-Oppenheimer approximation, a relativistic Lagrangian 
density which gives the same mean-field equations
of motion for a charmed or a bottom hypernucleus,  
in which the quasi-particles moving
in single-particle orbits are three-quark clusters with the quantum numbers
of a charmed baryon, a bottom baryon or a nucleon, 
when expanded to the same order in velocity, 
is given by QMC~\cite{Guichonf,Saitof,Tsushima_hyp}: 
\begin{eqnarray}
{\cal L}^{CHY}_{QMC} &=& {\cal L}_{QMC} + {\cal L}^C_{QMC}, 
\non \\
{\cal L}_{QMC} &=&  \overline{\psi}_N(\vec{r}) 
\left[ i \gamma \cdot \partial
- M_N^\star(\sigma) - (\, g_\omega \omega(\vec{r}) 
+ g_\rho \frac{\tau^N_3}{2} b(\vec{r}) 
+ \frac{e}{2} (1+\tau^N_3) A(\vec{r}) \,) \gamma_0 
\right] \psi_N(\vec{r}) \quad \non \\
  &-& \frac{1}{2}[ (\nabla \sigma(\vec{r}))^2 +
m_{\sigma}^2 \sigma(\vec{r})^2 ]
+ \frac{1}{2}[ (\nabla \omega(\vec{r}))^2 + m_{\omega}^2
\omega(\vec{r})^2 ] \non \\
 &+& \frac{1}{2}[ (\nabla b(\vec{r}))^2 + m_{\rho}^2 b(\vec{r})^2 ]
+ \frac{1}{2} (\nabla A(\vec{r}))^2, \non \\
{\cal L}^C_{QMC} &=& \sum_{C=\Lambda^+_c,\Lambda_b}     
\overline{\psi}_C(\vec{r}) 
\left[ i \gamma \cdot \partial
- M_C^\star(\sigma)
- (\, g^C_\omega \omega(\vec{r}) 
+ g^C_\rho I^C_3 b(\vec{r}) 
+ e Q_C A(\vec{r}) \,) \gamma_0 
\right] \psi_C(\vec{r}), \qquad \label{Lagrangian}
\end{eqnarray}
where $\psi_N(\vec{r})$ ($\psi_C(\vec{r})$)
and $b(\vec{r})$ are respectively the
nucleon (charmed and bottom baryon) and the $\rho$
meson (the time component in the third direction of
isospin) fields, while $m_\sigma$, $m_\omega$ and $m_{\rho}$ are
the masses of the $\sigma$, $\omega$ and $\rho$ meson fields.
$g_\omega$ and $g_{\rho}$ are the $\omega$-$N$ and $\rho$-$N$
coupling constants which are related to the corresponding
($u,d$)-quark-$\omega$, $g_\omega^q$, and
($u,d$)-quark-$\rho$, $g_\rho^q$, coupling constants as
$g_\omega = 3 g_\omega^q$ and
$g_\rho = g_\rho^q$~\cite{Guichonf,Saitof}.
Hereafter we will use notations for the quark flavors,
$q \equiv u,d$ and $Q \equiv s,c,b$.

In an approximation where the $\sigma$, $\omega$ and $\rho$ fields couple
only to the $u$ and $d$ quarks,  
the coupling constants in the charmed and bottom  baryons  
are obtained as $g^C_\omega = (n_q/3) g_\omega$, and
$g^C_\rho \equiv g_\rho = g_\rho^q$, with $n_q$ being the total number of
valence $u$ and $d$ (light) quarks in the baryon $C$. $I^C_3$ and $Q_C$
are the third component of the baryon isospin operator and its electric
charge in units of the proton charge, $e$, respectively.
The field dependent $\sigma$-$N$ and $\sigma$-$C$
coupling strengths predicted by the QMC model,
$g_\sigma(\sigma)$ and  $g^C_\sigma(\sigma)$,
related to the Lagrangian density, 
Eq.~(\ref{Lagrangian}), at the hadronic level are defined by:
\bg
M_N^\star(\sigma) &\equiv& M_N - g_\sigma(\sigma)
\sigma(\vec{r}) ,  \\
M_C^\star(\sigma) &\equiv& M_C - g^C_\sigma(\sigma)
\sigma(\vec{r}) , \label{coupny}
\en
where $M_N$ ($M_C$) is the free nucleon (charmed and bottom baryon) 
mass (masses).
Note that the dependence of these coupling strengths on the applied
scalar field must be calculated self-consistently within the quark
model~\cite{Guichonf,Saitof,Tsushima_hyp}. 
Hence, unlike quantum hadrodynamics (QHD)~\cite{QHD}, even though
$g^C_\sigma(\sigma) / g_\sigma(\sigma)$ may be
2/3 or 1/3 depending on the number of light quarks in the baryon 
in free space ($\sigma = 0$)\footnote{Strictly, this is true
only when the bag radii of nucleon and baryon $C$ are exactly the same
in the present model. See the last line in Eq.~(\ref{mit}).}, 
this will not necessarily  be the case in
nuclear matter.
More explicit expressions for $g^C_\sigma(\sigma)$
and $g_\sigma(\sigma)$ will be given later. From
the Lagrangian density, 
Eq.~(\ref{Lagrangian}), a set of
equations of motion for the charm 
or bottom hypernuclear system is obtained,
\begin{eqnarray}
& &[i\gamma \cdot \partial -M^\star_N(\sigma)-
(\, g_\omega \omega(\vec{r}) + g_\rho \frac{\tau^N_3}{2} b(\vec{r}) 
 + \frac{e}{2} (1+\tau^N_3) A(\vec{r}) \,) 
\gamma_0 ] \psi_N(\vec{r}) = 0, \label{eqdiracn}\\
& &[i\gamma \cdot \partial - M^\star_C(\sigma)-
(\, g^C_\omega \omega(\vec{r}) + g_\rho I^C_3 b(\vec{r}) 
+ e Q_C A(\vec{r}) \,) 
\gamma_0 ] \psi_C(\vec{r}) = 0, \label{eqdiracy}\\
& &(-\nabla^2_r+m^2_\sigma)\sigma(\vec{r}) = 
- [\frac{\partial M_N^\star(\sigma)}{\partial \sigma}]\rho_s(\vec{r})  
- [\frac{\partial M_C^\star(\sigma)}{\partial \sigma}]\rho^C_s(\vec{r}),
\non \\
& & \hspace{7.5em} \equiv g_\sigma C_N(\sigma) \rho_s(\vec{r})
    + g^C_\sigma C_C(\sigma) \rho^C_s(\vec{r}) , \label{eqsigma}\\
& &(-\nabla^2_r+m^2_\omega) \omega(\vec{r}) =
g_\omega \rho_B(\vec{r}) + g^C_\omega 
\rho^C_B(\vec{r}) ,\label{eqomega}\\
& &(-\nabla^2_r+m^2_\rho) b(\vec{r}) =
\frac{g_\rho}{2}\rho_3(\vec{r}) + g^C_\rho I^C_3 \rho^C_B(\vec{r}),  
 \label{eqrho}\\
& &(-\nabla^2_r) A(\vec{r}) = 
e \rho_p(\vec{r}) 
+ e Q_C \rho^C_B(\vec{r}) ,\label{eqcoulomb}
\end{eqnarray}
where, $\rho_s(\vec{r})$ ($\rho^C_s(\vec{r})$), $\rho_B(\vec{r})$
($\rho^C_B(\vec{r})$), $\rho_3(\vec{r})$ and
$\rho_p(\vec{r})$ are the scalar, baryon, third component of isovector,
and proton densities at the position $\vec{r}$ in
the charmed or bottom hypernuclei~\cite{Guichonf,Saitof,Tsushima_hyp}.
On the right hand side of Eq.~(\ref{eqsigma}),
$- [\frac{\partial M_N^\star(\sigma)}{\partial \sigma}] = 
g_\sigma C_N(\sigma)$ and
$- [\frac{\partial M_C^\star(\sigma)}{\partial \sigma}] = 
g^C_\sigma C_C(\sigma)$, where $g_\sigma \equiv g_\sigma (\sigma=0)$ and
$g^C_\sigma \equiv g^C_\sigma (\sigma=0)$,   
are a new, and characteristic feature of QMC 
beyond QHD~\cite{QHD,ruf,Jennings}. 
The effective mass for the charmed or bottom baryon $C$ is defined by,
\begin{equation}
\frac{\partial M_C^\star(\sigma)}{\partial \sigma}
= - n_q g_{\sigma}^q \int_{bag} d\vec{y} 
\ {\overline \psi}_q(\vec{y}) \psi_q(\vec{y})
\equiv - n_q g_{\sigma}^q S_C(\sigma) = - \frac{\partial}{\partial \sigma}
\left[ g^C_\sigma(\sigma) \sigma \right],
\label{gsigma}
\end{equation}
with the MIT bag model quantities~\cite{Guichon,Guichonf,Saitof,Tsushima_hyp},
\begin{eqnarray}
& &M_C^\star(\sigma) =
\sum_{j=q,Q}\frac{n_j\Omega^*_j -  z_C}{R_C^*}
+ \frac{4}{3}\pi ({R_C^*})^3 B ,\non \\
& &S_C(\sigma) = \frac{\Omega_q^*/2 
+ m_q^*R_C^*(\Omega_q^*-1)}
{\Omega_q^*(\Omega_q^*-1) 
+ m_q^*R_C^*/2},\non \\ 
& &
\Omega_q^* = \sqrt{x_q^2 + (R_C^* m_q^*)^2}, 
\quad \Omega_Q^* = \sqrt{x_Q^2 + (R_C^* m_Q)^2},\quad 
m_q^* = m_q - g_{\sigma}^q \sigma (\vec{r}), \non \\
& &C_C(\sigma) = S_C(\sigma)/S_C(0), \quad
g^C_{\sigma} \equiv n_q g_{\sigma}^q S_C(0) 
= \frac{n_q}{3} g_\sigma S_C(0)/S_N(0) 
\equiv \frac{n_q}{3}g_\sigma \Gamma_{C/N}. \label{mit}
\end{eqnarray}
Quantities for the nucleon are similarly obtained by replacing the 
indices, $C \to N$.
Here, $z_C$, $B$, $x_{q,Q}$, and $m_{q,Q}$ are the parameters
for the sum of the c.m.
and gluon fluctuation effects,
bag pressure, lowest eigenvalues for the quarks, $q$ or $Q$, respectively,
and the corresponding current quark masses.
$z_N$ and $B$ ($z_C$) are fixed by fitting the nucleon 
(charmed or bottom baryon) mass
in free space.
Concerning the sign of $m_q^*$ in (hyper)nucleus, 
it reflects nothing but the strength 
of the attractive scalar potential, 
and thus naive interpretation of the mass for a (physical) particle,      
which is positive, should not be applied.

The bag radii in-medium, $R_{N,C}^*$, are obtained
by the equilibrium condition\\
$d M_{N,C}^\star(\sigma)/{d R_{N,C}}|_{R_{N,C}=R_{N,C}^*} = 0$.
The bag parameters calculated and chosen for 
the present study in free space are,
$(z_N,\, z_\Lambda,z_{\Lambda^+_c},z_{\Lambda_b})$  
$= (3.295, 3.131, 1.766, -0.643)$, 
$(R_N, R_\Lambda, R_{\Lambda^+_C}, R_{\Lambda_b})$  
$= (0.800, 0.806, 0.846, 0.930)$ fm,
$B^{1/4} = 170$ MeV, $(m_{u,d},m_s,m_c,m_b)$ $= (5,250,1300,4200)$ MeV.
The parameters associated with the $u$, $d$ and $s$ quarks are the same as in
our previous investigations~\cite{Saitof,Tsushima_hyp}.
At the hadron level, the entire information on the quark dynamics is condensed
in $C_{N,C}(\sigma)$ of Eq.~(\ref{eqsigma}).
The parameters at the hadron level, which are already fixed by the study of
infinite nuclear matter and finite nuclei~\cite{Saitof},
are as follows: $m_\omega = 783$ MeV, $m_\rho = 770$ MeV, $m_\sigma = 418$ MeV,
$e^2/4\pi = 1/137.036$, $g^2_\sigma/4\pi = 3.12$, $g^2_\omega/4\pi = 5.31$
and $g^2_\rho/4\pi = 6.93$.

We briefly discuss about the spin-orbit force in QMC~\cite{Guichonf}. 
The origin of the spin orbit force for a composite nucleon moving
through scalar and vector fields which vary with position was explained
in detail in Ref.~\cite{Guichonf} -- c.f. sect. 3.2. The situation for the
$\Lambda$ and also for other hyperons are discussed in detail 
in Ref.~\cite{Tsushima_hyp}. 

In order to include the spin-orbit potential (approximately) 
correctly, e.g., for 
the $\Lambda^+_c$, 
we add perturbatively the correction due to the vector potential,
$ -\frac{2}{2 M^{\star 2}_{\Lambda^+_c} (\vec{r}) r}
\, \left( \frac{d}{dr} g^{\Lambda^+_c}_\omega \omega(\vec{r}) \right)
\vec{l}\cdot\vec{s}$,
to the single-particle energies obtained with the Dirac
equation, in the same way as that added in Ref.~\cite{Tsushima_hyp}.
This may correspond to a correct spin-orbit force which 
is calculated by the underlying quark 
model~\cite{Guichonf,Tsushima_hyp}:
\begin{equation}
V^{\Lambda^+_c}_{S.O.}(\vec{r}) \vec{l}\cdot\vec{s}
= - \frac{1}{2 M^{\star 2}_{\Lambda^+_c} (\vec{r}) r}
\, \left( \frac{d}{dr} [ M^\star_{\Lambda^+_c} (\vec{r})
+ g^{\Lambda^+_c}_\omega \omega(\vec{r}) ] \right) \vec{l}\cdot\vec{s},
\label{soQMC}
\end{equation}
since the Dirac equation at the hadronic level solved in usual QHD-type 
models leads to:
\begin{equation}
V^{\Lambda^+_c}_{S.O.}(\vec{r}) \vec{l}\cdot\vec{s}
= - \frac{1}{2 M^{\star 2}_{\Lambda^+_c} (\vec{r}) r}
\, \left( \frac{d}{dr} [ M^\star_{\Lambda^+_c} (\vec{r})
- g^{\Lambda^+_c}_\omega \omega(\vec{r}) ] \right) \vec{l}\cdot\vec{s},
\label{soQHD}
\end{equation}
which has the opposite sign for the vector potential,  
$g^{\Lambda^+_c}_\omega \omega(\vec{r})$.
The correction to the spin-orbit force, which appears naturally in the
QMC model, may also be modeled at the hadronic level of the Dirac equation by
adding a tensor interaction, motivated by the quark 
model~\cite{Jennings2,Cohen}.
Here, we should make a comment that, as was discussed 
by Dover and Gal~\cite{Gal} 
in detail, one boson exchange model with underlying (approximate) 
SU(3) symmetry in strong interaction, also leads to the weaker 
spin-orbit forces for the (strange) hyperon-nucleon ($YN$) 
than that for the nucleon-nucleon ($NN$).

However, in practice, because of its heavy mass 
($M^\star_{\Lambda^+_c}$), contribution to the single-particle energies 
from the spin-orbit potential with or without 
including the correction term, turned 
out to be even smaller than that for the 
$\Lambda$-hypernuclei, and further smaller for the $\Lambda_b$-hypernuclei.
Contribution from the spin-orbit potential with the correction term 
is typically of order $0.01$ MeV,   
and even for the largest case is $\cong 0.1$ MeV. 
This can be understood when one considers the limit, 
$M^\star_{\Lambda^+_c} \to \infty$
in Eq.~(\ref{soQMC}), where the quantity inside the square brackets varies   
smoothly from an order of hundred MeV to zero near the surface of 
the hypernucleus, and the derivative with respect to $r$ is finite.    
(See also Figs.~\ref{Capot} and~\ref{Pbpot}.)

\begin{figure}[hbt]
\begin{center}
\epsfig{file=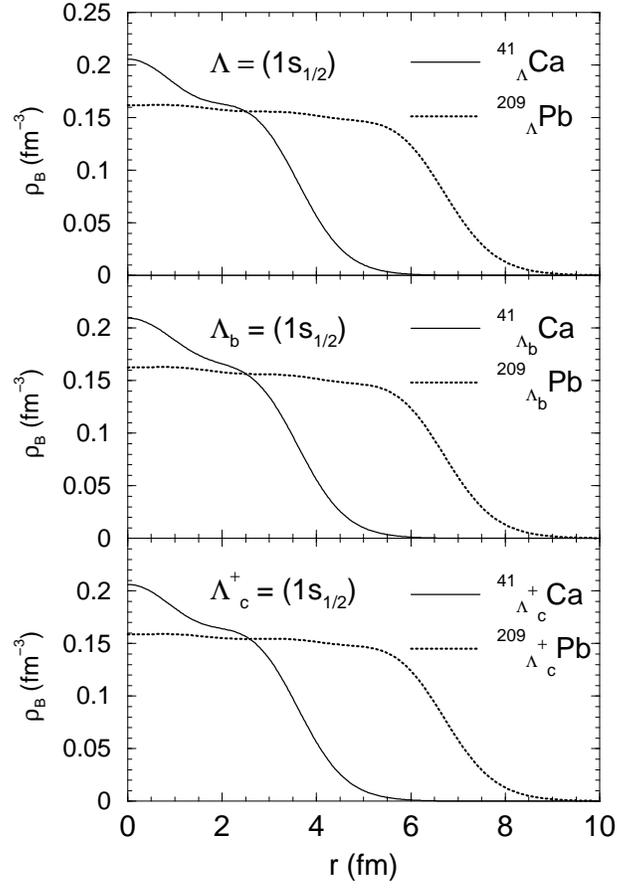,height=12cm,angle=-90}
\caption{
Total baryon density distributions in  
$^{41}_j$Ca and $^{209}_j$Pb ($j = \Lambda,\Lambda^+_c,\Lambda_b$),  
for $1s_{1/2}$ configuration for the $\Lambda, \Lambda^+_c$ and $\Lambda_b$.
\label{CaPbden}
}
\end{center}
\end{figure}

\begin{figure}[hbt]
\begin{center}
\epsfig{file=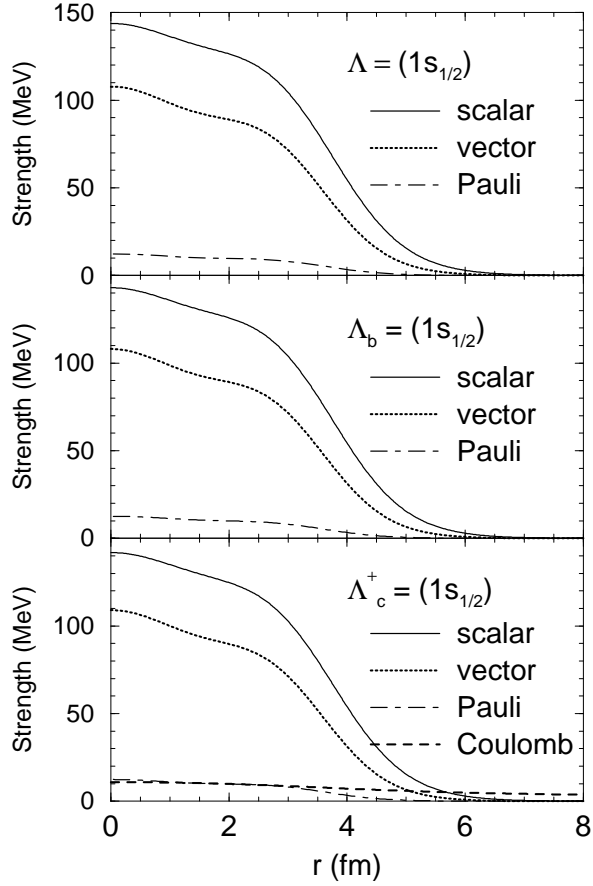,height=12cm,angle=-90}
\caption{
Potential strengths for $1s_{1/2}$ state felt by 
the $\Lambda,\Lambda^+_c$ and $\Lambda_b$ 
in $^{41}_j$Ca ($j = \Lambda,\Lambda^+_c,\Lambda_b$).
"Pauli" stands for the effective, repulsive, potential representing 
the Pauli blocking at the quark level plus
the $\Sigma_{c,b} N - \Lambda_{c,b} N$ channel coupling,
introduced at the baryon level
phenomenologically~\protect\cite{Tsushima_hyp}.
\label{Capot}
}
\end{center}
\end{figure}

\begin{figure}[hbt]
\begin{center}
\epsfig{file=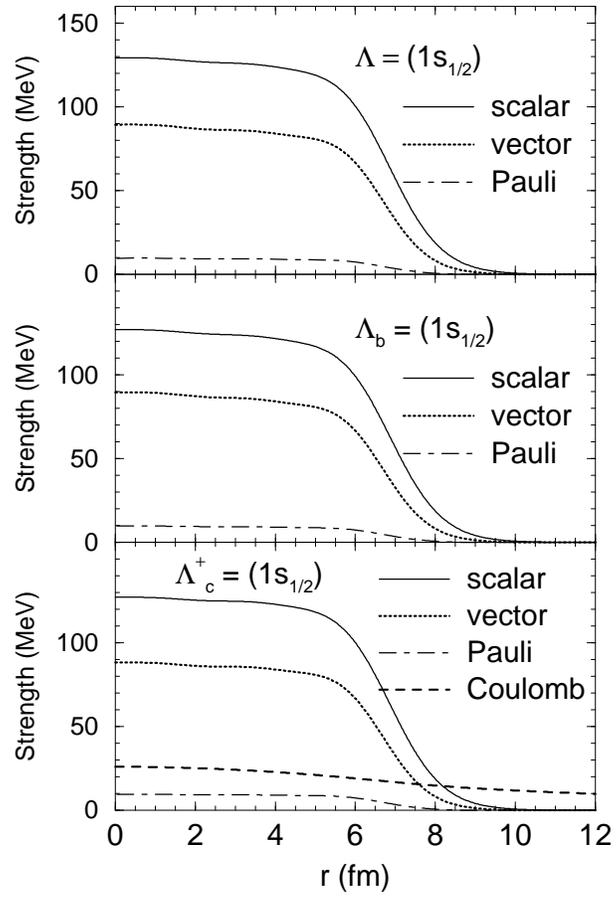,height=12cm,angle=-90}
\caption{
Potential strengths for $1s_{1/2}$ state felt by 
the $\Lambda,\Lambda^+_c$ and $\Lambda_b$ 
in $^{209}_j$Pb ($j = \Lambda,\Lambda^+_c,\Lambda_b$).
See also caption of Fig.~\protect\ref{Capot}.
\label{Pbpot}
}
\end{center}
\end{figure}

Now we discuss the results.
First, we show in Fig.~\ref{CaPbden} the total baryon density distributions 
in $^{41}_j$Ca and $^{209}_j$Pb ($j=\Lambda,\Lambda^+_c,\Lambda_b$), 
for $1s_{1/2}$ configuration in each hypernucleus. 
Note that because of the self-consistency, the total baryon density 
distributions are dependent on the configurations of 
the embedded particles. The total baryon density distributions are quite 
similar for the $\Lambda$-, $\Lambda^+_c$- and $\Lambda_b$-hypernuclei
multiplet which has the same baryon numbers, A, since the effect of 
$\Lambda,\Lambda^+_c$ and $\Lambda_b$ is $\cong 1/A$ for 
each hypernucleus.
Nevertheless, one notices that the 
$\Lambda_b$-hypernuclei density near the center 
is slightly higher than the corresponding 
$\Lambda$- and $\Lambda^+_c$-hypernuclei. 
This is because the $\Lambda_b$ is 
heavy and localized nearer the center, and 
contributes to the total baryon density there.
The baryon (probability) density distributions for the 
$\Lambda$, $\Lambda^+_c$ and $\Lambda_b$ in corresponding 
hypernuclei will be shown later.

Next, in Figs.~\ref{Capot} and~\ref{Pbpot}, we show the scalar and vector 
potentials felt by the $\Lambda$, $\Lambda^+_c$ and $\Lambda_b$ 
for $1s_{1/2}$ state 
in $^{41}_j$Ca and $^{209}_j$Pb ($j=\Lambda,\Lambda^+_c,\Lambda_b$), 
and the corresponding probability density distributions in Fig.~4. 
In Figs. 2 and 3 
"Pauli" stands for the effective, repulsive, potential representing 
the Pauli blocking at the quark level plus 
the $\Sigma_{c,b} N - \Lambda_{c,b} N$ channel coupling,  
introduced at the baryon level 
phenomenologically~\cite{Tsushima_hyp}.
For the $\Lambda^+_c$, the Coulomb potentials are also shown. 
As for the case of the nuclear matter~\cite{TK}, the scalar and vector 
potentials felt by these particles in hypernuclei multiplet 
which has the same baryon numbers are also quite similar. 
Thus, as far as the total baryon density distributions and 
the scalar and vector potentials are concerned, 
$\Lambda$-, $\Lambda^+_c$- and $\Lambda_b$-hypernuclei show quite similar 
features within the multiplet. 
However, as shown in Fig.~4, 
the wave functions obtained for $1s_{1/2}$ state are very different.
The $\Lambda^+_c$ baryon density distribution in $^{209}_{\Lambda^+_c}$Pb
is much more pushed away from the center
than that for the $\Lambda$ in $^{209}_\Lambda$Pb due to the Coulomb force.
On the contrary, the $\Lambda_b$ baryon density distributions
in $\Lambda_b$-hypernuclei are much larger near the origin than those for
the $\Lambda$ in the corresponding $\Lambda$-hypernuclei
due to its heavy mass.

Having obtained reasonable ideas about the potentials felt by  
$\Lambda,\Lambda^+_c$ and $\Lambda_b$, we show 
the calculated single-particle energies in 
Tables~\ref{table1} and~\ref{table2}.
Results for the $\Lambda$-hypernuclei are from Ref.~\cite{Tsushima_hyp}.
In these calculations, effective Pauli blocking,  
the effect of the $\Sigma_{c,b} N - \Lambda_{c,b} N$ 
channel coupling, and correction to the
spin-orbit force based on the underlying quark structure 
are included in the same way as adopted in Ref.~\cite{Tsushima_hyp}.
(However, recall that the negligibly small contribution from 
the correction terms for the spin-orbit 
force, and also contributions from the spin-orbit force itself.) 
Note that since the mass difference of the $\Lambda_c^+$ and $\Sigma_c$ 
is larger than that of the $\Lambda$ and $\Sigma$,   
and it is probably also true for the $\Lambda_b$ and $\Sigma_b$, 
we expect the effect of the channel coupling for the charmed and bottom 
hypernuclei is smaller than those for the 
strange hypernuclei, although the same parameters 
were used in the present calculation.
In addition, we searched for the single-particle states up to 
the same highest state as that of the core neutrons in each hypernucleus, 
since the deeper levels are usually easier to observe in experiment.

In Tables~\ref{table1} and~\ref{table2}, 
it is clear that the $\Lambda^+_c$ 
single-particle energy levels are higher 
than the corresponding levels for the $\Lambda$ and $\Lambda_b$. 
This is a consequence of the 
Coulomb force. This feature becomes stronger as 
the proton number in the core nucleus increases. 

Second, the level spacing for the $\Lambda_b$ single-particle energies 
is much smaller than that for the $\Lambda$ and $\Lambda^+_c$.
This may be ascribed to its heavy mass (or $M^\star_b$). 
In the Dirac equation for the $\Lambda_b$, 
the mass term dominates more than that of the term  
proportional to Dirac's $\kappa$, which classifies the states,  
or single-particle 
wave functions. (See Refs.~\cite{Saitof,Tsushima_hyp} for detail.)
This small level spacing would make it very difficult to  
distinguish the states in experiment, 
or to achieve such high resolution.
On the other hand, this may imply also many new phenomena. 
It will have a large probability to trap a $\Lambda_b$ among 
one of those many states, 
especially in heavy nucleus such as lead (Pb).
What are the consequences ? May be the  
$\Lambda_b$ weak decay happens inside a heavy nucleus with a very low 
probability ? Does it emit many photons when the $\Lambda_b$ gradually 
makes transitions from a deeper state to a shallower state ?
All these questions raise a flood of speculations.

To summarize, we have made a quantitative study for  
$\Lambda^+_c$- and $\Lambda_b$-hypernuclei in a quark-meson coupling 
model. We have solved a system of equations in a self-consistent 
approach for several finite nuclei with closed shell plus 
a hyperon ($\Lambda^+_c$ or $\Lambda_b$), embedding 
a $\Lambda^+_c$ or $\Lambda_b$ in the nucleus. 
Results are compared with those for the 
$\Lambda$-hypernuclei. It is shown that, although the scalar and 
vector potentials felt by the $\Lambda$, $\Lambda^+_c$- and $\Lambda_b$ 
are quite similar in corresponding hypernuclei multiplet which has the 
same baryon numbers, 
the single-particle wave functions, and single-particle energy level
spacings are quite different. 
For the $\Lambda^+_c$-hypernuclei, the Coulomb force  
plays a crucial role, and so does the heavy $\Lambda_b$ mass 
for the $\Lambda_b$-hypernuclei. 
It should be emphasized that we have used the values  
for the coupling constants of $\sigma$ 
(or $\sigma$-field dependent strength), $\omega$ and $\rho$ 
to the $\Lambda, \Lambda^+_c$ and $\Lambda_b$ determined 
automatically based on the underlying quark model, as for the 
nucleon and other baryons.
(Recall that the values for the vector $\omega$ fields to 
any baryons can be obtained by the number of light quarks in a baryon, 
but those for the $\sigma$ are different as 
shown in Eqs.~(\ref{gsigma}) and~(\ref{mit}).)
Phenomenology would determine ultimately if the coupling constants 
(strengths) determined by the underlying quark model 
actually work for $\Lambda^+_c$ and $\Lambda_b$ or not.
Although implications of the present results can be speculated 
a great deal, we would like to emphasize that,  
what we showed is that the $\Lambda^+_c$- and $\Lambda_b$-hypernuclei 
would exist in realistic experimental conditions.  
Experiments at facilities like JHF would provide further inputs to gain 
a better understanding of the interaction of $\Lambda^+_c$ 
and $\Lambda_b$ with the nuclear matter. 
Additional studies are needed to investigate the semi-leptonic
weak decay of $\Lambda^+_c$ and $\Lambda_b$ hyperons. 
The role of Pauli blocking and density in influencing the decay rates 
as compared to the free hyperons would be highly useful.
Such study can have an impact on the hadronization of the quark-gluon 
plasma and the transport of hadrons in nuclear matter of high density.
Will the high density lead to a slower decay and that a higher 
probability to survive its passage through the material ?
At present the study of the presence of 
$\Lambda^+_c$ and $\Lambda_b$ in finite nuclei is its infancy.
Careful investigations, both theoretical and experimental, would lead to 
a much better understanding of the role of heavy quarks in finite nuclei.

\vspace{1cm}
\noindent
Acknowledgment\\
The authors would like to thank Prof. A.W. Thomas for the hospitality 
at the CSSM, Adelaide, where this work was initiated.
K.T. acknowledges support and warm hospitality at University of 
Alberta, where this work was completed.
K.T. is supported by the Forschungszentrum-J\"{u}lich, 
contract No. 41445282 (COSY-058). The work of F.K. is 
supported by NSERCC. 


%
\newpage
%
\begin{table}[htbp]
\begin{center}
\caption{Single-particle energies (in MeV)
for $^{17}_j$O, $^{41}_j$Ca and $^{49}_j$Ca 
($j=\Lambda,\Lambda^+_c,\Lambda_b$). 
Single-particle energy levels are calculated up to the same highest states 
as that of the core neutrons. 
Results for the hypernuclei are taken 
from Ref.~\protect\cite{Tsushima_hyp}.
Experimental data for $\Lambda$-hypernuclei 
are taken from Ref.~\protect\cite{chr}, where 
spin-orbit splittings for $\Lambda$-hypernuclei 
are not well determined by the experiments.}
\label{table1}
\begin{tabular}[t]{c|cccc|cccc|ccc}
\hline 
&$^{16}_\Lambda$O  &$^{17}_\Lambda$O 
&$^{17}_{\Lambda^+_c}$O  &$^{17}_{\Lambda_b}$O
&$^{40}_\Lambda$Ca &$^{41}_\Lambda$Ca
&$^{41}_{\Lambda^+_c}$Ca &$^{41}_{\Lambda_b}$Ca
&$^{49}_\Lambda$Ca    &$^{49}_{\Lambda^+_c}$Ca &$^{49}_{\Lambda_b}$Ca\\
&(Exp.)& & & &(Exp.)& & & & & & \\
\hline
$1s_{1/2}$&-12.5&-14.1&-12.8&-19.6&-20.0&-19.5&-12.8&-23.0&-21.0&-14.3&-24.4\\
$1p_{3/2}$&-2.5 &-5.1 &-7.3 &-16.5&-12.0&-12.3&-9.2 &-20.9&-13.9&-10.6&-22.2\\
$1p_{1/2}$&($1p_{3/2}$)&-5.0&-7.3 &-16.5
&($1p_{3/2}$)&-12.3&-9.1&-20.9&-13.8&-10.6&-22.2\\ 
$1d_{5/2}$&     &     &     &     &     &-4.7 &-4.8 &-18.4&-6.5 &-6.5 &-19.5\\
$2s_{1/2}$&     &     &     &     &     &-3.5 &-3.4 &-17.4&-5.4 &-5.3 &-18.8\\
$1d_{3/2}$&     &     &     &     &     &-4.6 &-4.8 &-18.4&-6.4 &-6.4 &-19.5\\
$1f_{7/2}$&     &     &     &     &     &     &     &     &---  &-2.0 &-16.8\\
\end{tabular}
\end{center}
\end{table}
%
\newpage
%
%
\begin{table}[htbp]
\begin{center}
\caption{Single-particle energies (in MeV)
for $^{91}_j$Zr and $^{208}_j$Pb 
($j=\Lambda,\Lambda^+_c,\Lambda_b$).
Experimental data are taken from Ref.~\protect\cite{aji}.
See caption of Table~\protect\ref{table1} for other explanations.
}
\label{table2}
\begin{tabular}[t]{c|cccc|cccc}
\hline 
&$^{89}_\Lambda$Yb          &$^{91}_{\Lambda}$Zr
&$^{91}_{\Lambda^+_c}$Zr    &$^{91}_{\Lambda_b}$Zr
&$^{208}_\Lambda$Pb         &$^{209}_{\Lambda}$Pb
&$^{209}_{\Lambda^+_c}$Pb   &$^{209}_{\Lambda_b}$Pb\\
&(Exp.)& & & &(Exp.)& & & \\
\hline 
$1s_{1/2}$&-22.5&-23.9&-10.8&-25.7&-27.0&-27.0&-5.2 &-27.4\\
$1p_{3/2}$&-16.0&-18.4&-8.7 &-24.2&-22.0&-23.4&-4.1 &-26.6\\
$1p_{1/2}$&($1p_{3/2}$)&-18.4&-8.7 &-24.2
&($1p_{3/2}$)&-23.4&-4.0 &-26.6\\
$1d_{5/2}$&-9.0 &-12.3&-5.8 &-22.4&-17.0&-19.1&-2.4 &-25.4\\
$2s_{1/2}$&---  &-10.8&-3.9 &-21.6&---  &-17.6&---  &-24.7\\
$1d_{3/2}$&($1d_{5/2}$)&-12.3&-5.8 &-22.4
&($1d_{5/2}$)&-19.1&-2.4 &-25.4\\
$1f_{7/2}$&-2.0 &-5.9 &-2.4 &-20.4&-12.0&-14.4&---  &-24.1\\
$2p_{3/2}$&---  &-4.2 &---  &-19.5&---  &-12.4&---  &-23.2\\
$1f_{5/2}$&($1f_{7/2}$)&-5.8 &-2.4 &-20.4
&($1f_{7/2}$)&-14.3&---  &-24.1\\
$2p_{1/2}$&     &-4.1 &---  &-19.5&---  &-12.4&---  &-23.2\\
$1g_{9/2}$&     &---  &---  &-18.1&-7.0 &-9.3 &---  &-22.6\\
$1g_{7/2}$&     &     &     &     
&($1g_{9/2}$)&-9.2 &---  &-22.6\\
$1h_{11/2}$&    &     &     &     &     &-3.9 &---  &-21.0\\
$2d_{5/2}$&     &     &     &     &     &-7.0 &---  &-21.7\\
$2d_{3/2}$&     &     &     &     &     &-7.0 &---  &-21.7\\
$1h_{9/2}$&     &     &     &     &     &-3.8 &---  &-21.0\\
$3s_{1/2}$&     &     &     &     &     &-6.1 &---  &-21.3\\
$2f_{7/2}$&     &     &     &     &     &-1.7 &---  &-20.1\\
$3p_{3/2}$&     &     &     &     &     &-1.0 &---  &-19.6\\
$2f_{5/2}$&     &     &     &     &     &-1.7 &---  &-20.1\\
$3p_{1/2}$&     &     &     &     &     &-1.0 &---  &-19.6\\
$1i_{13/2}$&    &     &     &     &     &---  &---  &-19.3\\
\end{tabular}
\end{center}
\end{table}
%
%
\newpage
\newpage
\newpage
\newpage
\begin{figure}[hbt]
\begin{center}
\epsfig{file=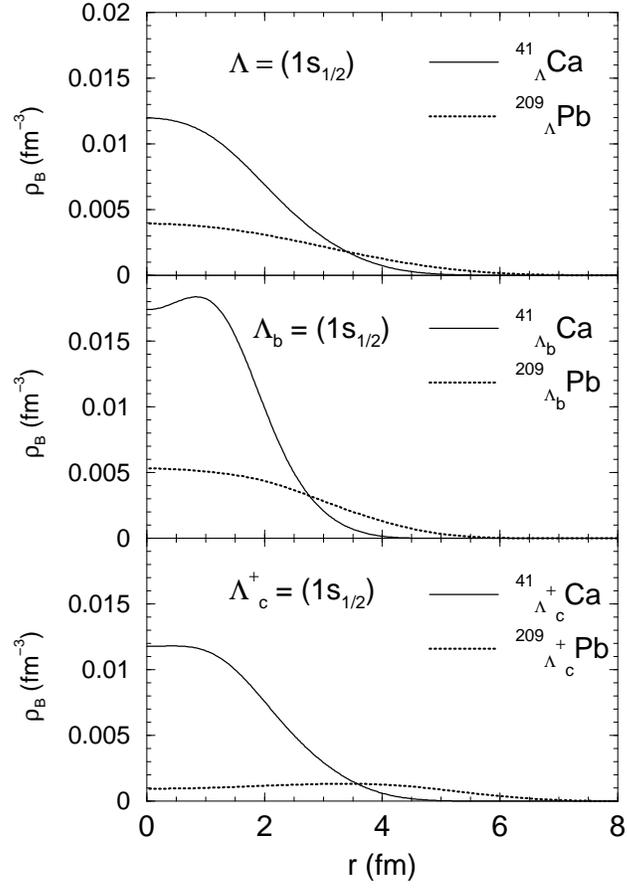,height=12cm,angle=-90}
\caption{$\Lambda,\Lambda^+_c$ and $\Lambda_b$ 
baryon (probability) density distributions for $1s_{1/2}$ state 
in $^{41}_j$Ca and 
$^{209}_j$Pb ($j = \Lambda,\Lambda^+_c,\Lambda_b$).
\label{HBdensity}
}
\end{center}
\end{figure}

%
\end{document}